# Interplay between emission wavelength and s-p splitting in MOCVD-grown InGaAs/GaAs quantum dots emitting above 1.3 μm

Paweł Podemski,[1,*] Anna Musiał,[1] Krzysztof Gawarecki,[2] Aleksander Maryński,[1] Przemysław Gontar,[1] Artem Bercha,[3] Witold A. Trzeciakowski,[3] Nicole Srocka,[4] Tobias Heuser,[4] David Quandt,[4] André Strittmatter,[4,†] Sven Rodt,[4] Stephan Reitzenstein[4] and Grzegorz Sęk[1]

[1]*Laboratory for Optical Spectroscopy of Nanostructures, Department of Experimental Physics, Wrocław University of Science and Technology, 50-370 Wrocław, Poland.*
[2]*Department of Theoretical Physics, Wrocław University of Science and Technology, 50-370 Wrocław, Poland.*
[3]*Institute of High Pressure Physics, Polish Academy of Sciences, 01-142 Warsaw, Poland.*
[4]*Institute of Solid State Physics, Technische Universität Berlin, 10623 Berlin, Germany.*

E-mail: pawel.podemski@pwr.edu.pl

The electronic structure of strain-engineered single InGaAs/GaAs quantum dots emitting in the telecommunication O band is probed experimentally by photoluminescence excitation spectroscopy. Observed resonances can be attributed to p-shell states of individual quantum dots. The determined energy difference between s-shell and p-shell shows an inverse dependence on the emission energy. The experimental data are compared with the results of confined states calculations, where the impact of the size and composition in the investigated structures is simulated within the 8-band **k·p** model. On this basis, the experimental observation is attributed mainly to changes in indium content within individual quantum dots, indicating a way of engineering and selecting a desired quantum dot, whose electronic structure is the most suitable for a given nanophotonic application.

---

[†] Present address: Institute of Experimental Physics, Otto von Guericke University Magdeburg, D-39106 Magdeburg, Germany.

Security of data transmission has become an important issue in information technology. Today, most information is stored and exchanged digitally, with most transfers being secured by public-key cryptography. However, the development of quantum computing has shown disturbingly easy ways to overcome classical cryptography security, which assumes limited computer performance. In particular, Shor's algorithm has demonstrated that number factorisation (and thus the data decryption) can be performed in computation time shorter by a polynomial factor.[1] This has led to intensive research focusing on truly secure communication – quantum cryptography.[2] For practical realisation of device-independent quantum key distribution and quantum communication via optical fibres, single-photon sources operating at telecommunication wavelengths are key building blocks.[3]

One of the most promising systems for realising single-photon sources based on non-classical light emitters are semiconductor quantum dots (QDs),[4] which have been shown to work well at telecommunication wavelengths, e.g., in the low dispersion and low loss O band.[5–7] Moreover, for this spectral range optical-fibre-compatible single-photon sources based on quantum dots have been demonstrated.[8-10] Further optimisation of the optical properties of such sources is required to increase the collection efficiency and the fabrication yield, for instance by decreasing the QD areal density and by deterministic integration of QDs into optical microcavities or other photonic structures increasing the extraction of the emitted photons.[11] To obtain efficient emission in the telecom O band InGaAs QDs on GaAs substrate (typically emitting below 1 μm) can be used in conjunction with an InGaAs strain-reducing layer (SRL) shifting the emission to the desired 1.3 μm.[12-16] The incorporation of an additional layer in the vicinity of QDs during the growth process influences the local strain and therefore their size and composition, as well as the depth of the confining potential, modifying the electronic structure expected for common and well-established InGaAs/GaAs QDs emitting below 1 μm (i.e., those without additional strain-reducing layers). Therefore, to fully understand and use these as quantum light sources it is essential to identify their electronic structure. In particular, the highest purity of single-photon emission of InGaAs QDs can be obtained using p-shell excitation scheme[17] and this requires precise knowledge of the confined exciton excited states spectrum of individual QDs, including the p-shell state energies. Moreover, for the design of these sources it would be beneficial to have the possibility to affect or tailor both, the QD ground state energy (i.e. emission wavelength) and the energetic distance from the ground state to the excited state, preferably separated by at least thermal energy at a given operation temperature from the emitting quantum dot ground state, to provide high thermal stability and spectral isolation. Therefore, it is crucial to identify the main determinants of the electronic structure of these application-relevant epitaxial nanostructures, that are especially important for the envisioned nanophotonic applications, such as efficient sources of single photons or entangled photon pairs emitting at telecommunication wavelengths compatible with existing fibre networks. Although some attempts with quasi-resonant excitation have been performed for quantum-optical experiments,[9] there is no systematic study on the interrelation between the emission wavelength and energy separation between the ground and excited states in single, O band-emitting InGaAs/GaAs QDs with strain-reducing layer or on identification of the decisive structural properties.

In this report we determine the energetic splitting between s-shell and p-shell states (s-p splitting) in single InGaAs/GaAs QDs emitting in the O band, and we show how the s-p splitting can be influenced by the QD composition, size and the In content in the SRL. The investigated sample was grown by metalorganic chemical vapour deposition (MOCVD) on a GaAs substrate with 23 pairs of GaAs/Al$_{0.9}$Ga$_{0.1}$As layers forming a distributed Bragg reflector (DBR), providing increased emission collection efficiency from the layer of QDs. InGaAs quantum dots (with nominal indium content of 0.75) were formed during Stranski-Krastanov self-assembled growth followed by capping with 4 nm of In$_{0.2}$Ga$_{0.8}$As, creating a strain-reducing layer, resulting in indium concentration and QD sizes favouring 1.3 μm emission. Finally, a 630-nm-thick GaAs layer was grown, from which an array of microlenses (with no specific QDs preselection) was produced by means of 3D electron-beam lithography.[18,19] In this etching process QDs outside lenses were removed, leaving areas of QDs with a base diameter of 3.6 μm. The areal density of these QDs was of the order of $10^9$ cm$^{-2}$ with their base diameter of 30 nm, and an average height of 6-8 nm.[9,20] A reference sample was also used with an ensemble of similar QDs with slightly different indium composition and with no DBR below the dot layer to prevent overlapping of the DBR characteristics with the spectrally broad emission of a QD ensemble in the spectroscopic studies. To identify excited states in these quantum dots single-dot photoluminescence excitation (PLE) spectroscopy adapted to the spectral range above 1 μm was used.[21] QDs were excited by a self-made continuous wave external-cavity tunable laser followed by a short-focal-length monochromator and shortpass filters to provide a clean excitation laser line. For nonresonant excitation a 639 nm semiconductor laser was used. The QD sample was mounted in a continuous-flow microscopy cryostat providing temperatures down to 5 K. To excite single QDs and collect their emission, an achromatic objective with 0.4 numerical aperture was applied, offering a laser spot diameter on the sample surface of single micrometres, which was small enough to excite selectively single microlenses. QD emission spectra were detected by a nitrogen-cooled linear array InGaAs sensor coupled to a monochromator.

Standard photoluminescence (PL) spectrum for the unpatterned reference sample with the whole ensemble of InGaAs QDs is presented in Fig. 1(a). This shows the spectral range of QD emission for nonresonant excitation (639 nm; 1.940 eV) with the excitation power density of 4 kW/cm$^2$. Two maxima related to the radiative recombination in QDs which exhibit the typical state filling effect with increasing excitation power (not shown here) can be observed. Thus, the left maximum corresponds to the QD ground state (s-shell) emission, while the presence of the second maximum is attributed to the recombination in a higher (p-shell) QD state and not to a multimodal size distribution.[22] The reference sample with an ensemble of similar QDs has been studied in detail in Ref. 20. The energy difference between the QD maxima is approximately 70 meV, indicating the expected splitting energy between s-shell and p-shell exciton states within a single InGaAs QD. This value is relatively large as compared to conventional InGaAs QDs emitting below 1 µm,[17,23] and should provide high thermal stability of the ground state emission. On the high energy slope of the second maximum, a fingerprint of an even higher quantum dot state, separated by approx. 130 meV from the ground state, can be observed – confirming a rather deep confining potential of the investigated QDs. Figure 1(b) presents an exemplary photoluminescence spectrum from a single QD excited quasi-resonantly (1242 nm; 0.998 eV; i.e. into the expected p-shell state, based on the high-excitation PL of the QD ensemble) with the excitation power density of 1 kW/cm$^2$, with several sharp emission lines observed in the O band spectral range.

An example of a complete PLE map from single InGaAs QDs is presented in Fig. 2(a). This shows an evolution of the emission lines from Fig. 1(b) as the excitation energy is tuned in the range of the expected p-shell state energies. Indeed, at certain excitation energy values several of the lines exhibit an enhancement of the emission intensity, suggesting an increase of the absorption when at the resonance between the excitation energy and a higher energy state within a QD. Figure 2(b) shows single QD PLE spectra, extracted from the map in Fig. 2(a), for two QD emission lines: 1336.9 nm (red) and 1342.5 nm (violet). The maximum in the red PLE spectrum appears 70 meV above the ground state emission energy and the peak in the violet PLE spectrum is located at the energy difference of 74.5 meV. These resonances exhibit the lowest measured energy difference to the emission energy, which is a fingerprint of excitation into the first excited state (p-shell). Both these values are in agreement with the energy difference between the excited and ground quantum dot states expected from measurements on an ensemble of similar QDs.[20] The PLE maxima are rather broad (~2 meV), which may be related to the relatively large excitation laser linewidth (~0.7 meV),[21] or could result from a complex energy structure (i.e. dense ladder of states) of the QD excited states.[20] The energy difference of the observed PLE resonances (preliminarily identified as related to the p-shell absorption) changes with the energy of the emission line. The energy difference between the PLE maxima and the ground state emission (assigned provisionally as "s-p splitting") of numerous single InGaAs QDs studied is presented in Fig. 3(a) as a function of the QD ground state energy, where the excitation power density was kept at the level of 1 kW/cm$^2$ for all cases. There is a clear dependence observed: QDs emitting at higher energies exhibit a significant decrease of "s-p splitting" energy (from 80 meV down to 60 meV). Both energies should indeed be related, but the expected dependence is opposite in typical QD systems. The ground state energy in self-assembled QDs is usually a direct result of the size distribution of QDs within an ensemble, which is mainly influenced by the QD height providing the strongest quantization, and the smaller the dot the higher its ground state energy. However, a decrease of the QD size should at the same time increase the s-p splitting due to stronger energy levels quantization – this is opposite to the behaviour observed for the InGaAs QDs investigated in this study. It is worth noting that the s-p splitting mainly depends on the overall single-particle shell energy structure of a QD and less on the effects related to Coulomb interactions and various carrier combinations (excitonic complexes) confined within a QD. The influence of Coulomb interactions on the s-p splitting is typically one order of magnitude weaker and is up to several millielectronvolts in our calculations. Thus, the exact character of the excitonic complexes related to individual emission lines is of less importance in the present study.

To confirm the identification of the observed PLE resonances as the p-shell absorption and to explain the s-p splitting dependence, the single InGaAs QD energy structure was modelled and its evolution with QD size and material composition was evaluated. The strain distribution in the system was calculated within a continuous elasticity approach,[24] using the second-order deformation potentials[25] with a piezoelectric field based on strain-induced polarization up to the second order in strain tensor elements.[26] In the calculations an improved set of deformation potentials was used (for InAs: $a_c$ = -4.78; $a_c^{(2a)}$ = -3.40; $a_c^{(2b)}$ = 18.1; $a_v$ = 1.260; $a_v^{(2a)}$ = -1.04; $a_v^{(2b)}$ = 0.925; $b_v$ = -1.763; $b_v^{(2a)}$ = -3.98; $b_v^{(2b)}$ = -6.94; for GaAs: $a_c$ = -6.79; $a_c^{(2a)}$ = -2.71; $a_c^{(2b)}$ = 24.7; $a_v$ = 1.88; $a_v^{(2a)}$ = -2.62; $a_v^{(2b)}$ = 4.29; $b_v$ = -1.84; $b_v^{(2a)}$ = -3.58; $b_v^{(2b)}$ = -9.5; all in eV). A gradient of indium distribution inside the QD as suggested by structural data (not shown here) was also included, where indium was concentrated at the centre of a QD – see Ref. 27. Single-particle electron and hole states were calculated within the 8-band $\mathbf{k}\cdot\mathbf{p}$ model,[28] while the excitonic states were obtained within a configuration interaction approach. The numerical values of the s-p splitting were calculated from the energy difference between the lowest bright configuration with significant p-type contributions and the bright exciton s-type state. More details of the calculations and material parameters are given in Ref. 29. For realistic QD parameters (e.g., diameter of 30 nm, height of 6 nm, lens shape of QD) the calculated QD ground state energy is 0.92 eV with the s-p splitting energy of approximately 80 meV, corresponding well to the measured PLE resonances. Thus, confirming the identification of the observed experimental maxima. Next, the influence of the QD size on the s-p splitting energy was simulated and the result is shown in Fig. 3(b) – the QD size is changed relatively to the abovementioned dimensions (see the red curve) with the size multiplier indicated in the figure - all QD dimensions are changed simultaneously, which is typically expected, in the first approximation, for such III-V self-assembled QDs.[13,30,31] A decrease in QD size shifts the ground state energy to higher values and increases the s-p splitting, which is in contrast to the experimentally observed dependence. The other parameter expected to alter the electronic structure in InGaAs QDs is the indium content. Its influence on the s-p splitting energy was calculated for the strain-reducing layer and QDs, independently, and is also presented in Fig. 3(b). Here, the In content corresponds to the average In amount within the QD and to In content in SRL, where no composition gradient is assumed. These results show similar trends as observed in the experiment – higher ground state energy is associated with lower s-p splitting values. Thus, the dominant factor responsible for s-shell and p-shell separation in InGaAs QDs covered by an SRL is not their size but the indium content. Changing the SRL content exclusively gives a significantly weaker dependence than the experimental one, and would be able to cover only a small part of the experimentally obtained values and only for unrealistically broad In content range fluctuations from 5% to 35%. Therefore, this factor must be of secondary importance. On the contrary, modifying the QD composition gives the s-p splitting vs emission energy dependence much closer to the experimental data, for still realistic In contents.

The average indium content in the simulated QD was changed from 0.75 down to 0.61. For the quantum dot emitting at 0.92 eV, the average In content in the dot (obtained within the simulated gradient distribution) is 0.73, which is slightly lower than the nominal indium content of 0.75, but still very close (e.g., the accuracy of experimentally determined In content in Ref. 27 was ±0.13). In the simulations the maximum In content in the gradient was 1.0, which corresponds to the average In content of 0.75. This value constitutes the upper limit of the possible In content for the assumed In distribution within the QD. To reproduce the experimental data, QDs with an average indium content down to 0.61 were also simulated, which suggests a lower In content in the investigated QDs emitting at higher energies. It is worth noting that an increase of the average indium content (i.e. InAs amount within the InGaAs alloy) leads to lower ground state energy, but at the same time the effective mass is reduced, resulting in the larger separation between s-shell and p-shell states. The change of the indium content also influences the lattice mismatch between InGaAs and GaAs, and the resulting strain field has an impact on the QD energy structure – this effect was taken into account in the calculations. For the investigated QDs all the simulated parameters (i.e. QD size; QD composition; SRL composition) do change within the ensemble simultaneously. Therefore, the absolute energy values from the simulations do not correspond to the experimental s-p slitting values precisely, but support (or do not) the observed trends.

In general, one can expect that s-p splitting energy is also influenced by the QD aspect ratio. For comparison, the effect of varying the QD height is also shown in Fig. 3(b) as a grey dashed line where the indium composition gradient is also rescaled accordingly in the QD height direction, while the base QD size remains constant. It is worth noting that the ground state energy is increasing for higher QDs, which shows that its changes are driven predominantly by the strain and not by the spatial confinement. Although the obtained s-p splitting dependence could, in principle, reflect the experimental observation, we consider this scenario as unlikely to occur. Firstly, as mentioned above, during the growth of self-assembled QDs the aspect ratio does not usually change significantly, i.e. the increase in the QD diameter is followed by a simultaneous increase of the dot height.[13,30,31] Secondly, in order to cover the experimental range of emission energies one would need to assume a very broad change of the QDs height (with a constant QD base diameter), up to exceeding 10 nm, which is already at the limit for plastic relaxation[32] and deteriorates QD radiative properties. Some changes in the QDs' height cannot be ruled out from our considerations, and these perhaps contribute to the obtained inverse s-p splitting vs emission energy dependence. However, taking into account all the limitations, the latter seems to be driven by the QD composition predominantly.

It is worth commenting that in previous studies higher indium content in InGaAs/GaAs QDs was linked to smaller nanostructure size, however those QDs were grown by molecular beam epitaxy and were emitting below 1 μm, so this tendency does not have to transfer directly to other QDs in the same material system.[33] In our structures the s-p splitting can be influenced by QD size and composition change in opposite ways, as related to the QD ground state energy – see the red and green lines in Fig. 3(b). This demonstrates the possibility of mutual tuning of QD ground and excited states energies, offering an additional degree of freedom in quantum nanodevices engineering.

In Fig. 3(b), three points are shown for simulated QD emitting at 0.92 eV, but with different s-p splitting values ranging from 81 to 67 meV. This result was obtained for different pairs of QD size and indium average content within the QD volume – triangle: size 1.0, $In_x$ 0.73; circle: size 1.1, $In_x$ 0.71; rhombus: size 1.2, $In_x$ 0.70. In general, it is possible to obtain different s-p splitting values for QDs emitting at the same wavelength, when both the parameters (QD size and composition) are controlled. However, such straightforward control is beyond what is achievable technologically during the growth of QD structures, as it is impossible to keep the QD size constant and change the composition only. Nevertheless, the QD ground state and s-p splitting can be controlled on the whole ensemble level by a mutual change of QD size and indium composition, for instance by overgrowth of the quantum dot layer with InGaAs or thermally induced diffusion. From such an ensemble, individual QDs providing ground state and s-p splitting energy values, required for a given application, can be selected. Furthermore, when looking at the calculated s-p splitting for electron and hole single particle states, it is shared approximately 35:65 between the valence and conduction bands, respectively, corresponding to 19-26 meV for holes and 35-48 meV for electrons (depending on the specific QD), with approximately 6 meV provided in addition by the Coulomb interaction. This offers an energetic separation large enough to prevent any significant thermally-induced losses, resulting in the overall thermal stability of devices based on such InGaAs QDs, especially that current solutions still require cryogenic temperatures to operate, where the obtained separation of 80 meV between s-shell and p-shell states is sufficiently high, e.g., to allow for stand-alone operation in a compact Stirling cryocooler with base temperatures in the range of 30-40 K.[34]

In general, tailoring the s-p splitting could also have an impact on other parameters important for a quantum source efficiency, e.g., signal-to-background ratio, single-photon emission or fine-structure splitting. This has been partly considered in Refs. 9 and 27, but requires further study to obtain an accurate statistical information (e.g. on single-photon emission), so this is beyond the scope of this report.

In summary, we have studied InGaAs/GaAs QDs capped with a strain-reducing layer, providing QD emission redshift to the telecom O band. Single QD photoluminescence excitation spectroscopy has allowed to determine the energy splitting between s-shell and p-shell in individual QDs, revealing a reduced splitting for dots emitting at higher energies. Supported by theoretical modelling, this behaviour has been associated to be dominated by a varying indium content in different QDs within the ensemble. A secondary effect which could result in observed dependence is the aspect ratio. However, its distribution within a QD ensemble is expected to be narrow and therefore the influence on the s-p splitting energy is rather small. The ability to influence the interplay between the ground state emission and the s-p splitting energy, indicates an additional degree of freedom in the design and growth of QDs for O band spectral range applications, such as telecom single-photon sources with p-shell quasi-resonant optical pumping.[9] In particular, this allows to obtain an improved spectral isolation of the ground state transition, e.g. to increase the activation energy for carriers' escape via higher energy states in a QD and provide high temperature stability. The investigated structures were grown by MOCVD, which is also of practical importance, since this technology offers lower production costs, and hence is better suited for large-scale device fabrication than molecular beam epitaxy.

**Acknowledgments**
We acknowledge financial support from: the National Science Centre (Poland) within project No. 2014/15/D/ST3/00813; FI-SEQUR project jointly financed by the European Regional Development Fund (EFRE) of the European Union in the framework of the programme to promote research, innovation and technologies (Pro FIT) in Germany, and the National Centre for Research and Development in Poland within the 2[nd] Poland-Berlin Photonics Programme, grant No. 2/POLBER-2/2016 (project value PLN 2,089,498); the Polish National Agency for Academic Exchange; and the "Quantum dot-based indistinguishable and entangled photon sources at telecom wavelengths" project, carried out within the HOMING programme of the Foundation for Polish Science co-financed by the European Union under the European Regional Development Fund.

**References**

[1]P. Shor, SIAM J. Comput. **26**, 1484 (1997).

[2]H.-K. Lo, M. Curty, and K. Tamaki, Nat. Photon. **8**, 595 (2014).

[3]N. Sangouard and H. Zbinden, J. Mod. Opt. **59**, 1458 (2012).

[4]*Nano-Optics and Nanophotonics: Quantum Dots for Quantum Information Technologies*, ed. P. Michler (Springer, 2017).

[5]F. Klopf, R. Krebs, J.P. Reithmaier, and A. Forchel, IEEE Photonics Technology Letters **13**, 764 (2001).

[6]M. Sartison, L. Engel, S. Kolatschek, F. Olbrich, C. Nawrath, S. Hepp, M. Jetter, P. Michler, and S.L. Portalupi, Appl. Phys. Lett. **113**, 032103 (2018).

[7]A. Rantamäki, G.S. Sokolovskii, S.A. Blokhin, V.V. Dudelev, K.K. Soboleva, M.A. Bobrov, A.G. Kuzmenkov, A.P. Vasil'ev, A.G. Gladyshev, N.A. Maleev, V.M. Ustinov, and O. Okhotnikov, Opt. Lett., OL **40**, 3400 (2015).

[8]M.B. Ward, O.Z. Karimov, D.C. Unitt, Z.L. Yuan, P. See, D.G. Gevaux, A.J. Shields, P. Atkinson, and D.A. Ritchie, Appl. Phys. Lett. **86**, 201111 (2005).

[9]Ł. Dusanowski, P. Holewa, A. Maryński, A. Musiał, T. Heuser, N. Srocka, D. Quandt, A. Strittmatter, S. Rodt, J. Misiewicz, S. Reitzenstein, and G. Sęk, Opt. Express **25**, 31122 (2017).

[10]J. Kettler, M. Paul, F. Olbrich, K. Zeuner, M. Jetter, and P. Michler, Appl. Phys. B **122**, 48 (2016).

[11]P. Schneider, N. Srocka, S. Rodt, L. Zschiedrich, S. Reitzenstein, and Sven Burger, Opt. Express **26**, 8479 (2018).

[12]E. Goldmann, M. Paul, F.F. Krause, K. Müller, J. Kettler, T. Mehrtens, A. Rosenauer, M. Jetter, P. Michler, and F. Jahnke, Appl. Phys. Lett. **105**, 152102 (2014).

[13]F. Guffarth, R. Heitz, A. Schliwa, O. Stier, N. N. Ledentsov, A. R. Kovsh, V. M. Ustinov, and D. Bimberg, Phys. Rev. B **64**, 085305 (2001).

[14]B. Alloing, C. Zinoni, V. Zwiller, L. H. Li, C. Monat, M. Gobet, G. Buchs, A. Fiore, E. Pelucchi, and E. Kapon, Appl. Phys. Lett. **86**, 101908 (2005).

[15]A.E. Zhukov, A.R. Kovsh, N.A. Maleev, S.S. Mikhrin, V.M. Ustinov, A.F. Tsatsul'nikov, M.V. Maximov, B.V. Volovik, D.A. Bedarev, Yu.M. Shernyakov, P.S. Kop'ev, Zh.I. Alferov, N.N. Ledentsov, and D. Bimberg, Appl. Phys. Lett. **75**, 1926 (1999).

[16]L. Seravalli, M. Minelli, P. Frigeri, P. Allegri, V. Avanzini, and S. Franchi, Appl. Phys. Lett. **82**, 2341 (2003).

[17]P. Ester, L. Lackmann, S. Michaelis de Vasconcellos, M.C. Hübner, A. Zrenner, and M. Bichler, Appl. Phys. Lett. **91**, 111110 (2007).

[18]M. Gschrey, A. Thoma, P. Schnauber, M. Seifried, R. Schmidt, B. Wohlfeil, L. Krüger, J. -H. Schulze, T. Heindel, S. Burger, F. Schmidt, A. Strittmatter, S. Rodt, and S. Reitzenstein, Nat. Commun. **6**, 7662 (2015).

[19]N. Srocka, A. Musiał, P.-I. Schneider, P. Mrowiński, P. Holewa, S. Burger, D. Quandt, A. Strittmatter, S. Rodt, S. Reitzenstein, and G. Sęk, AIP Advances **8**, 085205 (2018).

[20]A. Maryński, P. Mrowiński, K. Ryczko, P. Podemski, K. Gawarecki, A. Musiał, J. Misiewicz, D. Quandt, A. Strittmatter, S. Rodt, S. Reitzenstein, and G. Sęk,

Acta Phys. Pol. A **132**, 386 (2017).

[21]P. Podemski, A. Maryński, P. Wyborski, A. Bercha, W. Trzeciakowski, and G. Sęk, Journal of Luminescence **212**, 300 (2019).

[22]U. W. Pohl, K. Pötschke, A. Schliwa, F. Guffarth, D. Bimberg, N. D. Zakharov, P. Werner, M. B. Lifshits, V. A. Shchukin, and D. E. Jesson, Phys. Rev. B **72**, 245332 (2005).

[23]A. Zrenner, F. Findeis, E. Beham, M. Markmann, G. Böhm, G. Abstreiter, Physica E **9**, 114 (2001).

[24]C. Pryor, J. Kim, L.W. Wang, A.J. Williamson, and A. Zunger, J. Appl. Phys. **83**, 2548 (1998).

[25]K. Gawarecki and M. Zieliński, Phys. Rev. B **100**, 155409 (2019).

[26]G. Bester, A. Zunger, X. Wu, and D. Vanderbilt, Phys. Rev. B **74**, 081305 (2006).

[27]P. Mrowiński, A. Musiał, K. Gawarecki, Ł. Dusanowski, T. Heuser, N. Srocka, D. Quandt, A. Strittmatter, S. Rodt, S. Reitzenstein, and G. Sęk, Phys. Rev. B **100**, 115310 (2019).

[28]T.B. Bahder, Phys. Rev. B **41**, 11992 (1990).

[29]K. Gawarecki, Phys. Rev. B **97**, 235408 (2018).

[30]D. Litvinov, H. Blank, R. Schneider, D. Gerthsen, T. Vallaitis, J. Leuthold, T. Passow, A. Grau, H. Kalt, C. Klingshirn, and M. Hetterich, J. Appl. Phys. **103**, 083532 (2008).

[31]A. Sauerwald, T. Kümmell, G. Bacher, A. Somers, R. Schwertberger, J. P. Reithmaier, and A. Forchel, Appl. Phys. Lett. **86**, 253112 (2005).

[32]J. F. Chen, Y. C. Lin, C. H. Chiang, Ross C. C. Chen, Y. F. Chen, Y. H. Wu, and L. Chang, Journal of Applied Physics **111**, 013709 (2012).

[33]A. Musiał, P. Gold, J. Andrzejewski, A. Löffler, J. Misiewicz, S. Höfling, A. Forchel, M. Kamp, G. Sęk, and S. Reitzenstein, Phys. Rev. B **90**, 045430 (2014).

[34]A. Schlehahn, S. Fischbach, R. Schmidt, A. Kaganskiy, A. Strittmatter, S. Rodt, T. Heindel, and S. Reitzenstein, Sci. Rep. **8**, 1340 (2018).

# Figure Captions

**Fig. 1.** (a) Photoluminescence spectrum from a reference sample with an ensemble of InGaAs/GaAs QDs obtained under non-resonant excitation conditions (639 nm; 1.940 eV) with estimated energy differences between consecutive QD states.
(b) Single QD photoluminescence spectrum for quasi-resonant excitation (1242 nm; 0.998 eV) of InGaAs/GaAs QDs. Both spectra were recorded at 5 K.

**Fig. 2.** (a) Single QD photoluminescence excitation map in the spectral region of excited states in InGaAs/GaAs QDs. (b) Single QD photoluminescence excitation spectra for two emission lines from the map (red: 1336.9 nm, violet: 1342.5 nm). The measurements were performed at 5 K.

**Fig. 3.** (a) InGaAs/GaAs QDs s-p splitting energies, derived from photoluminescence excitation measurements on many single QDs, as a function of their ground state energy (the line acts as a guide to the eye). (b) 8-band **k·p** simulations of InGaAs/GaAs QD for varying QD size (expressed by a size multiplier as described in the text), average indium composition in the QD, indium composition in the SRL and QD height (in nanometers). The resulting s-p splitting is displayed as a function of the QD ground state energy. The three data points at the energy of 0.92 eV correspond to different s-p splitting values, obtained from calculations for different sets of QD parameters (triangle: QD size 1.0, QD $In_x$ 0.73; circle: QD size 1.1, QD $In_x$ 0.71; rhombus: QD size 1.2, QD $In_x$ 0.70).

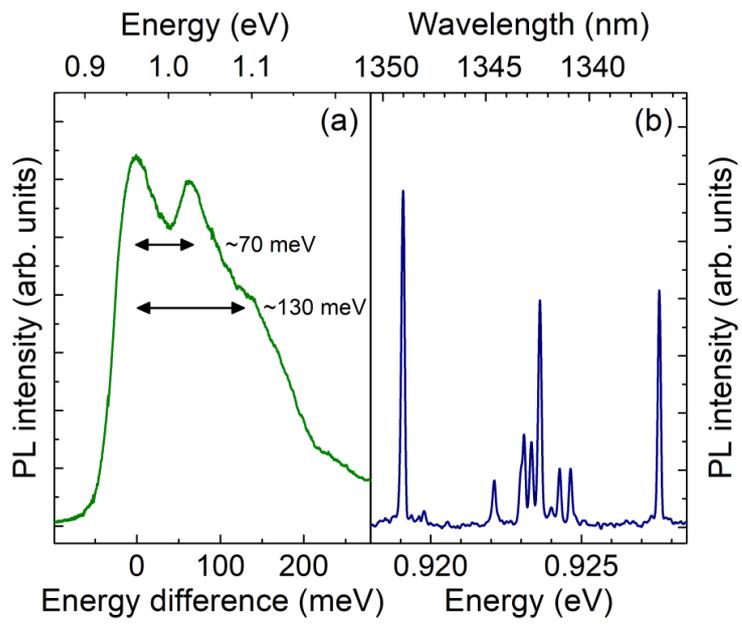

Fig. 1.

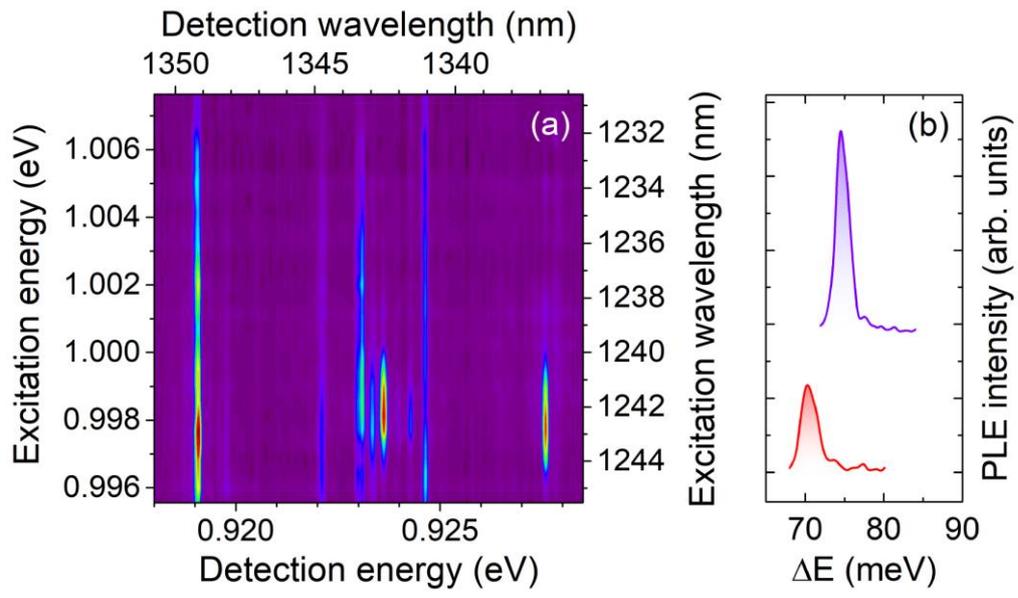

Fig. 2.

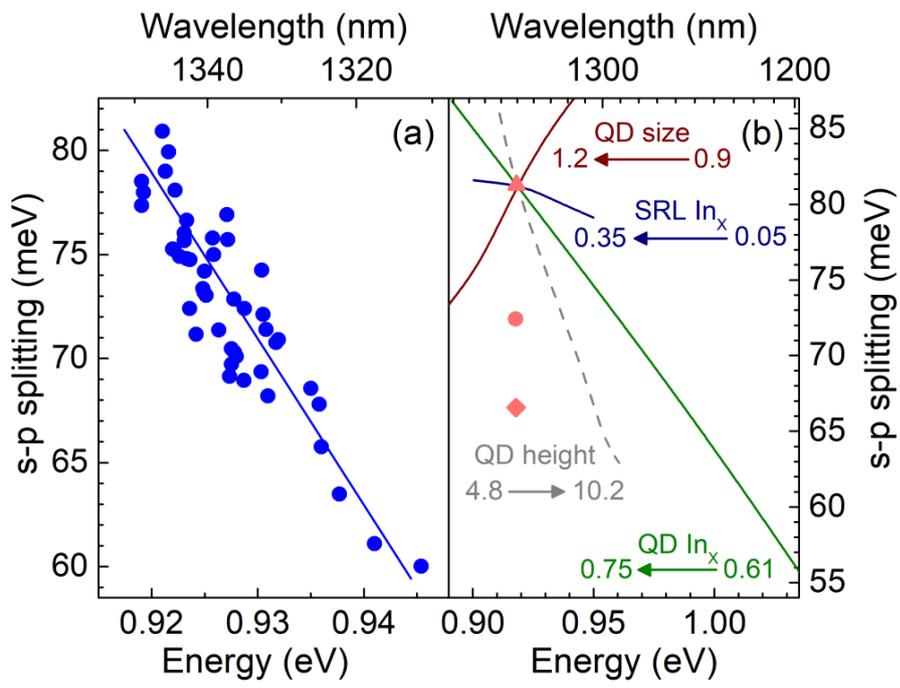

Fig. 3.